\documentclass{article}
\usepackage{spconf,amsmath,graphicx}
\usepackage{balance,subcaption,xcolor}

\usepackage{ctable}
\usepackage{float}
\usepackage{hyperref}
\usepackage{cite}
\usepackage{balance}

\title{Explaining deep learning models for spoofing and \\ deepfake detection with SHapley Additive exPlanations}

\name{Wanying Ge, Jose Patino, Massimiliano Todisco and Nicholas Evans \thanks{The first author is supported by the TReSPAsS-ETN project funded by the European Union Horizon 2020 research and innovation programme under the Marie Skłodowska-Curie grant agreement No.\ 860813.  The second author is supported by the ExTENSoR project funded by the French Agence Nationale de la Recherche (ANR).}}
\address{EURECOM, Sophia Antipolis, France}

\begin{document}
\ninept
\maketitle
\begin{abstract}

    Substantial progress in spoofing and deepfake detection has been made in recent years.  Nonetheless, the community has yet to make notable inroads in providing an explanation for how a classifier produces its output.  The dominance of black box spoofing detection solutions is at further odds with the drive toward trustworthy, explainable artificial intelligence.  This paper describes our use of SHapley Additive exPlanations (SHAP) to gain new insights in spoofing detection.  We demonstrate use of the tool in revealing unexpected classifier behaviour, the artefacts that contribute most to classifier outputs and differences in the behaviour of competing spoofing detection models. The tool is both efficient and flexible, being readily applicable to a host of different architecture models in addition to related, different applications.  All results reported in the paper are reproducible using open-source software. 

\end{abstract}
\begin{keywords}
spoofing, presentation attack detection, explainability, Shapley additive explanations, ASVspoof
\end{keywords}

\section{Introduction}
\label{sec:intro}
    Having begun over two decades ago~\cite{lindberg99_eurospeech, masuko99_eurospeech}, research in spoofing or presentation attack detection for automatic speaker verification and speech deepfake detection has gained traction only in the last six years~\cite{asvspoof2021}. Judging by the estimates of performance reported in the literature, typically measured through spoofing detection metrics such as the minimum tandem detection cost function~\cite{kinnunen2018tdcf} or the equal error rate, a great deal of progress has been achieved in recent years.

    While the community has been largely successful in developing effective detection solutions, it has yet to make substantial inroads to properly understanding and shedding light upon the artefacts used by spoofing countermeasures to distinguish between bona fide and spoofed speech.  This is especially the case for so-called {\em black-box} solutions which typify probably all current, state-of-the-art solutions to spoofing and deepfake detection~\cite{li21o_cRes2Net, tak21_gat,zhang21da_cas, hua2021TSSDNet, Luo2021Capsule}.

    The growing trend toward explainable artificial intelligence (xAI)~\cite{arrieta2020_xai} has not only the potential to produce machine learning models that provide some justification for scores or decisions, i.e.\ in the general vein of improving trustworthiness, but also to help researchers learn more about the problem in hand as well as the characteristics, behaviour, strengths and weaknesses of different solutions.  xAI approaches can also help reveal paths for future research and, ultimately, to expedite the development of better performing, more reliable and efficient machine learning solutions.
    
    We have hence set out to explore explainability in anti-spoofing.  Our goals are to better understand what artefacts are being captured or discarded by different solutions based upon different input features, different machine learning solutions or different components of ensemble systems~\cite{lavrentyeva2019stc_lcnn, wang2021comparative}. We are also hopeful of understanding why some spoofing attacks are more difficult to detect than others, e.g.\ the infamous A17 attack contained within the ASVspoof 2019 logical access (LA) database~\cite{asvspoof2019}.  We hope too to understand why solutions that operate directly upon raw waveform inputs~\cite{ hua2021TSSDNet,tak2021rawnet2,ge21_raw_pc_darts} can perform better than systems that operate upon hand-crafted spectro-temporal representations~\cite{wang2021comparative, chen2020generalization}, but worse for others, e.g.\ the A08 attack~\cite{ge21_raw_pc_darts}.  Additional motivation comes from the numerous reports in the literature which show that some solutions apply greater attention to non-speech intervals than to speech intervals~\cite{Bhusan_2017, cldnn, muller21_silence}.  Here we hope to better understand whether or not these issues are evidence of database design shortcomings and/or whether they point towards issues relating to the behaviour of certain, specific countermeasure solutions.  In case of the former, we hope that studies of explainability might contribute to the development of tools to assist with database quality control.

    This paper describes our exploration of SHapley Additive exPlanations (SHAP)~\cite{SHAP} to explain the behaviour of deep neural network solutions to spoofing and speech deepfake detection. While there are some related works~\cite{becker2018interpreting, muckenhirn19_visualiseCNN, sivasankaran21_SHAP_SE}, to the best of our knowledge, it is the first application of SHAP to the problem of spoofing detection. We describe SHAP and report examples of its application to utterances in the ASVspoof 2019 LA database. We demonstrate what can be learned from such tools and advocate for their broader adoption.  All results presented in this paper are fully reproducible using open-source codes.

\section{Related work}
    
    Several related explainability studies have been reported previously. Using an approach based upon the attenuation of distinct spectral components,~\cite{tak20_odyssey} shows that artefacts indicative of different spoofing attacks are located within different sub-band intervals, and hence that they can be detected more reliably with front-ends that emphasise the same frequency range. The use of Gradient-weighted Class Activation Mapping (Grad-CAM)~\cite{selvaraju2017_grad} to explain spoofing classifier behaviour is reported in~\cite{halpern20_odyssey}.  It is applied to generate a binary saliency map for the network input layer.  Input audio is then reconstructed using spectrograms masked with the binary saliency map.  Listening experiments show that the model uses the buzziness and rhythmic quality of speech sounds to distinguish between bona fide and spoofed speech. A study of replay detection~\cite{Bhusan_2017} shows the impact of different replay attack configurations upon detection performance. The use of Local Interpretable Model-agnostic Explanations (LIME)~\cite{ribeiro2016_LIME} to generate both temporal and spectral explanations of  model prediction behaviour for voice replay detection is reported in~\cite{Bhusan_2018_SLIME}. The input speech spectrogram is first segmented into a number of temporal or spectral segments, before LIME is applied to learn their relative importance through experiments with and without their use for modelling. These works show that non-speech intervals can provide discriminative information for spoofing detection. Related work in~\cite{muller21_silence} shows that the duration of non-speech intervals in a synthetic speech and converted voice detection task can also be indicative of whether an utterance is bona fide or spoofed.

    Drawing from concepts in cooperative game theory~\cite{Shapley53} whereby payoffs are distributed fairly to each player based on their individual contributions, SHapley Additive exPlanations (SHAP), introduced in~\cite{SHAP}, offers a more elegant and powerful approach to explainability.  SHAP values reflect the influence of particular features to a classifier output. The work in~\cite{sivasankaran21_SHAP_SE} reports the use of DeepSHAP~\cite{SHAP} to help explain the behaviour of speech enhancement models. SHAP values, estimates of Shapley values~\cite{Shapley53} derived using DeepSHAP, are used to determine regions of the input feature, expressed in the form of spectrograms, that impact most upon on the model output. Results show that better performing models tend to rely more on information within speech-dominated spectro-temporal intervals. SHAP provides a unified and theoretically grounded approach to explain the relative importance of particular features to classifier outputs. We have sought to apply SHAP analysis to help explain the behaviour of different spoofing detection models. We provide a brief overview of SHAP before describing its application to speech data and then spoofing detection.

\section{SHapley Additive exPlanations (SHAP)}
    
    SHAP values can be both positive and negative and reflect the relative (un)importance of a particular feature to a classifier output.  Given a prediction function (i.e., a model) $f(x)$ and a feature subset $S \subseteq F$, where $F$ is the full set of features, the SHAP value
    $\phi_{i}$ is derived using a pair of models, one learned with the inclusion of feature $i$, the other without. For an arbitrary input $x_{S}$, the predictions derived using both models are compared according to:
    \begin{equation} 
    \delta_{i}(S) = f_{S \cup \{i\}}(x_{S \cup \{i\}}) - f_{S}(x_{S})
    \label{equ:diff}
    \end{equation}
   where $f_{S \cup \{i\}}$ is the model trained with the feature subset supplemented with~$i$ and $f_{S}$ is the model trained on the same subset without the inclusion of~$i$. Eq.~(\ref{equ:diff}) is computed for all possible subsets $S \subseteq F \setminus \{i\}$ (subsets $S$ not containing $i$) in order to gauge the impact of withholding $i$ from the pool of features used for model training. The SHAP value is then given by: 
    \begin{equation}
        \label{eq:shapley}
    \phi_{i} = \sum_{S \subseteq F \setminus \{i\}} \frac{|S|!~(|F| - |S| -1)!}{|F|!}\delta_{i}(S) 
    \end{equation}
    which is a normalised average over the different permutations of $S \subseteq F \setminus \{i\}$, where $|S|$ and $|F|$ are the number of features in $S$ and $F$.
   
    By way of a simple, intuitive example, we consider an image (later a speech spectrogram) with a large number of pixels (the features, later short-time spectro-temporal magnitude estimates).  Consider the selection of one pixel $i$ in the image.  If the model predictions obtained by the model when it is trained with or without the inclusion of $i$ do not differ, then pixel $i$ bears little, specific relevance to the output.  On the other hand, differences between the two predictions would suggest that pixel $i$ is comparatively informative and bears a stronger influence upon the model output. 
    
    In the case that $f(\mathbf{x})$ is a complex model, such as a deep neural network, the calculation of $\phi_{i}$ according to Eq.~(\ref{eq:shapley}) is computationally prohibitive; the model must be trained twice for each feature subset~$S$. An alternative, more efficient approach is then needed.
    The input $\mathbf{x}$ is first simplified to take the form $\mathbf{x}'=\{x'_{1}, ..., x'_{D}\}$, where $x'_{i}\in\{0,1\}$ implies either the absence ($0$) or presence ($1$) of the corresponding feature in $\mathbf{x}$, and where $D$ is the feature dimension. An explanation model $g(\mathbf{x}')$ is then used as an approximation of $f(\mathbf{x})$:
    \begin{equation}
         \label{eq:SHAP}
        f(\mathbf{x})\approx g(\mathbf{x}') = \phi_{0}+ \sum_{i=1}^{D}\phi_{i}x'_{i} 
    \end{equation}
    where $\phi_{0}=f(h_x(\mathbf{0}))$ (all-zero input) and where $h_{x}$ is a mapping function that converts $\mathbf{x}'$ to $\mathbf{x}$, i.e., $\mathbf{x} = h(\mathbf{x}')$. The model output is hence approximated by the sum of SHAP values corresponding to the features for which $x'_i=1$. $g(\mathbf{x}')$ is then trained to approximate the original network output $f(\mathbf{x})$ and the coefficients $\phi_{i}$ of the model $g(\mathbf{x}')$ are used in place of true SHAP values~\cite{SHAP, molnar_2020}. 
    
    Even with these approximations, the calculation of SHAP values for deep neural networks remains computationally challenging. DeepSHAP~\cite{SHAP} provides an efficient method to estimate SHAP values for deep models. With assumptions of feature independence and model linearity, DeepSHAP approximates absent features with expected values.  This approach mitigates the need for repetitive model retraining in Eqs.~(\ref{equ:diff}) and~(\ref{eq:shapley}). SHAP values for simple network components (linear, max pooling, activation) are then estimated. The definition in Eq.~(\ref{eq:SHAP}) connects SHAP with DeepLIFT~\cite{deeplift}, an additive feature attribution method. DeepLIFT multipliers can then be passed via backpropagation to estimate SHAP values at the model input level.  Full details can be found in~\cite{SHAP, molnar_2020, lundberg2018consistent}.

\section{Application to speech data}
        \begin{figure}[!h]
     \centering
     \begin{subfigure}[b]{0.45\textwidth}
         \centering
     \includegraphics[width=\textwidth]{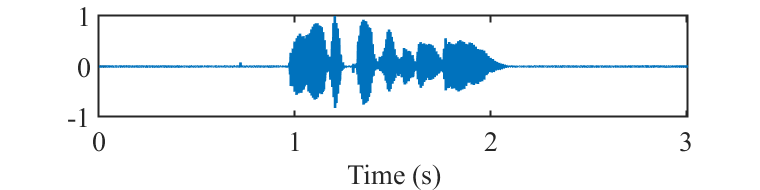}
        \caption{waveform}
         \label{fig:waveform}
     \end{subfigure}
     \hfill
     \begin{subfigure}[b]{0.45\textwidth}
         \centering
        \includegraphics[width=\textwidth]{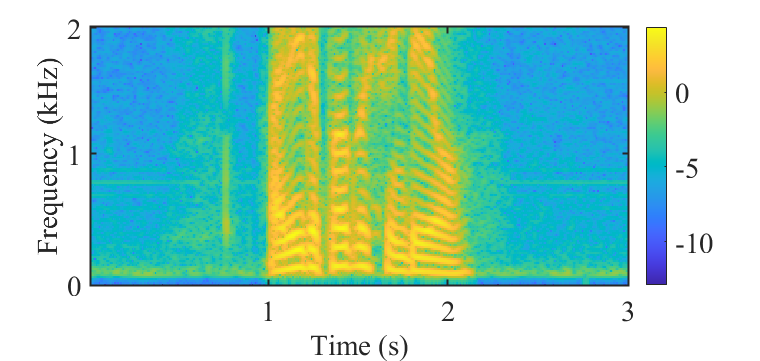}
         \caption{Log-scaled STFT spectrogram}
         \label{fig:spec}
     \end{subfigure}
     \hfill
     \begin{subfigure}[b]{0.45\textwidth}
         \centering
        \includegraphics[width=\textwidth]{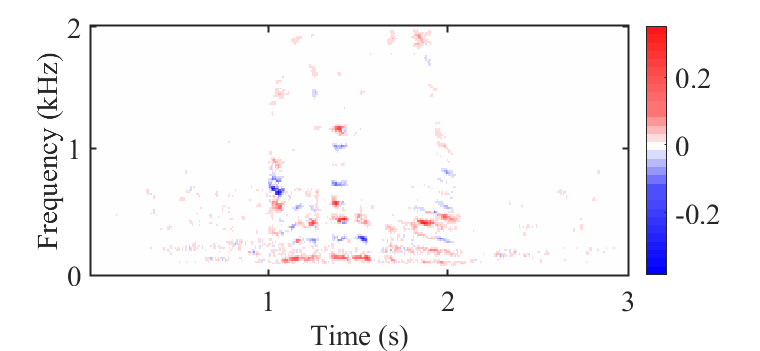}
         \caption{SHAP values for bona fide class}
         \label{fig:shap_bonafide}
     \end{subfigure}
     \hfill
     \begin{subfigure}[b]{0.45\textwidth}
         \centering
         \includegraphics[width=\textwidth]{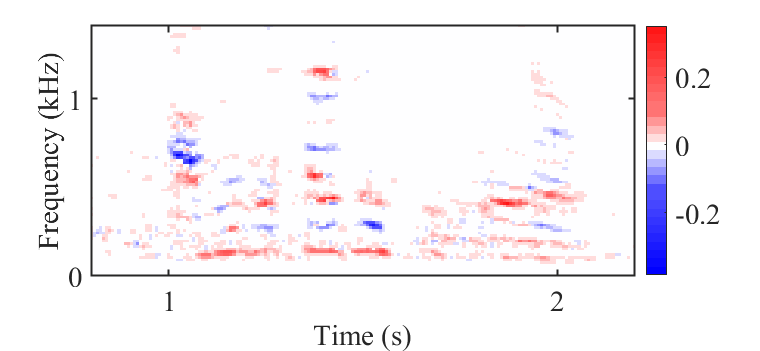}
         \caption{A zoomed version of (c)}
         \label{fig:shap_zoom}
     \end{subfigure}
        \caption{Illustrations of SHAP analysis for the `LA\_E\_3757378' bona fide utterance contained within the ASVspoof 2019 logical access evaluation dataset.}
        \label{fig:visulisation}
\end{figure}
    
    We describe how DeepSHAP can be applied to speech data and an arbitrary binary classifier. A time-series speech waveform $\mathbf{x}(t)$ sampled at 16~kHz is first transformed into a short-term spectro-temporal decomposition using the short-time Fourier transform (STFT) giving a spectrogram $\mathbf{X}(m, n)$, where $m$ is the spectral bin and where $n$ is the frame index.  An example time waveform and spectrogram are shown in Fig.~\ref{fig:waveform} and~\ref{fig:spec}. DeepSHAP is then applied treating a given (pre-trained) classifier (here a spoofing detection system) or network model as $f(\cdot)$ in Eq.~(\ref{eq:shapley}) and each spectro-temporal bin in $\mathbf{X}(m, n)$ individually for every pair ($m$,$n$) as a feature, akin to $i$ in Eq.~(\ref{equ:diff}), and the whole spectrogram as the full feature set $F$. For a given classifier, SHAP values $\phi_{i}$ in Eq.~(\ref{eq:shapley}) are calculated to determine the relative contribution of each point $(m, n)$ in $\mathbf{X}$ to the classifier output. For a binary classifier (later spoofing detection), a pair of $\phi_{i}$ are obtained, each one representing support for one of the two classes (later bona fide and spoof classes).

    SHAP values $\phi_{(m,n)}$
    can then be visualised in a similar manner to the spectrogram $\mathbf{X}$.  This process can be applied individually to any input $\mathbf{x}(t)$. An example for an arbitrary speech signal (here a bona fide speech utterance) is shown in Fig.~\ref{fig:shap_bonafide}. 
    It shows the degree to which each spectro-temporal bin contributes to the classifier output. Darker red points indicate the spectro-temporal bins which lend stronger support for the positive class (here bona fide).  In contrast, darker blue points indicate greater support for the negative class (here, spoofed speech).
    
    SHAP visualisations such as that in Fig.~\ref{fig:shap_bonafide} can be sparse, indicating that only few spectro-temporal bins contribute to the classifier output.  A comparison of the time waveform in Fig.~\ref{fig:waveform} and the SHAP values in Fig.~\ref{fig:shap_bonafide} shows that this particular classifier essentially ignores information contained in non-speech regions, focusing instead upon the speech interval between approximately~1 and 2~seconds and, furthermore between frequencies mostly below 1.5~kHz.  
    
    A second visualisation focusing on this specific region is displayed in Fig.~\ref{fig:shap_zoom}.  Ignoring for now whether or not the SHAP values are positive or negative, it exhibits a high degree of correlation to the fundamental frequency and harmonics in the spectrogram, indicating the focus of the classifier on these same components.  Last, while the presence of dark blue traces in Fig.~\ref{fig:shap_zoom} indicate components of the spectrogram which favour the negative class, the overall dominance of red colours (though not all {\em dark} red) indicate a greater support for the positive class (the classifier output correctly indicates bona fide speech).
    Plots of SHAP values such as those shown in Fig.~\ref{fig:shap_bonafide} are not easily visualised without the use of dilation operations or some other such smoothing operations which distort the results.  While they offer interesting insights, we need more easily visualised means with which to explore results.
    
\section{Explainability for spoofing detection}
    
    In the remainder of this paper we describe our use of DeepSHAP to help explain the behaviour of spoofing detection systems.  We show a number of illustrative examples for which the input utterances, all drawn from the ASVspoof 2019 LA database~\cite{asvspoof2019}, are chosen specially to demonstrate the potential insights which can be gained.  Given the difficulty in visualising true SHAP values, in the following we present average temporal or spectral results.  Given our focus on spoofing detection, we present results for both bona fide and spoofed utterances and the temporal or spectral regions which favour either bona fide or spoofed classes. Results hence reflect where, either in time or frequency, the model has learned to focus attention and hence help to explain its behaviour in terms of how the model responds to a particular utterance.

\subsection{Spoofing detection system}

    While the goal of this work is to demonstrate how DeepSHAP can be applied to the spoofing detection task, rather than an explainability study for a particular classifier, results are nonetheless specific to the latter.  We used two  different classifiers in this work. Source code is available for both and can be used with the open-source DeepSHAP toolkit\footnote{\href{https://github.com/slundberg/shap}{https://github.com/slundberg/shap}} to reproduce our results. The first model is our own partially connected differentiable architecture search (PC-DARTS) model described in~\cite{ge2021pc-darts}.\footnote{\href{https://github.com/eurecom-asp/pc-darts-anti-spoofing}{https://github.com/eurecom-asp/pc-darts-anti-spoofing}}  The second is the 2D-Res-TSSDNet model presented in~\cite{hua2021TSSDNet}.\footnote{\href{https://github.com/ghuawhu/end-to-end-synthetic-speech-detection}{https://github.com/ghuawhu/end-to-end-synthetic-speech-detection}} Both are built upon the same base concept of ResNets~\cite{he2016_resnet}, though the convolutional block architectures of the PC-DARTS model are optimised (searched) automatically rather than manually.  The models operate upon log-scaled power spectrograms extracted from 4-second utterances (PC-DARTS) and magnitude spectrograms extracted from 6-second utterances (2D-Res-TSSDNet), both of which are formed in the usual way from the concatenation or truncation of source data, and using 64~ms Hamming windows with a 8~ms shift and a 1024-point FFT. Both systems are competitive with other state-of-the-art, single systems~\cite{hua2021TSSDNet,ge2021pc-darts}. The expected value of a given input feature is computed from 20 different utterances of the same class (bona fide or the particular spoofing attack). 

\subsection{Classification using non-speech intervals}
\begin{figure}[!ht]
     \centering
     \begin{subfigure}[b]{0.45\textwidth}
         \centering
        \includegraphics[width=\textwidth]{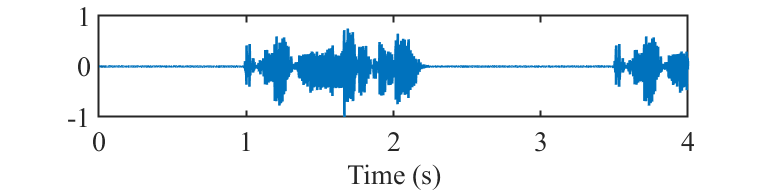} \caption{waveform}
         \label{fig:1832578}
     \end{subfigure}

     \begin{subfigure}[b]{0.45\textwidth}
         \centering
        \includegraphics[width=\textwidth]{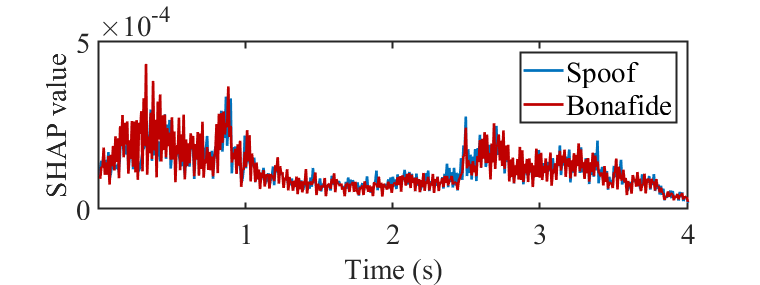} \caption{averaged positive SHAP values in time}
         \label{fig:pcdarts_shap_1832578}
     \end{subfigure}
        \caption{SHAP values for the `LA\_E\_1832578' utterance and the PC-DARTS model.} 
        \label{fig:pcdarts_1832578}
\end{figure}
    
    Fig.~\ref{fig:pcdarts_1832578} shows the results of SHAP analysis for the `LA\_E\_1832578' utterance and the PC-DARTS classifier.  The plot shows the time waveform (a) and the temporal variation in SHAP values averaged across the full spectrum (b).  This first example shows that the classifier has learned to focus predominantly upon non-speech intervals.  The support in speech intervals for either class is comparatively lower. These observations are unexpected; it is assumed a priori that spoofed speech detection systems should operate upon {\em speech}.  This observation corroborates the findings in~\cite{Bhusan_2017}, and also~\cite{muller21_silence} which shows that reliable bona fide/spoof decisions might even be inferred from the {\em length} of the non-speech interval.

\subsection{Sub-band artefacts}
\begin{figure}[!t]
         \centering
        \includegraphics[width=0.9\linewidth]{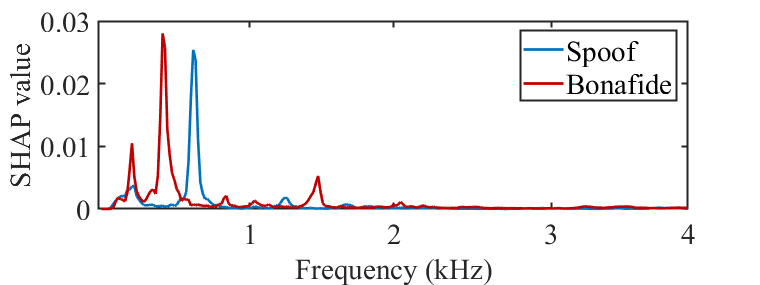}
        \caption{SHAP values for the `LA\_E\_2634822' utterance and the 2D-Res-TSSDNet model.}
        \label{fig:tssdnet_2634822}
\end{figure}
    Fig.~\ref{fig:tssdnet_2634822} shows time-averaged SHAP values against frequency for the spoofed `LA\_E\_2634822' utterance and the 2D-Res-TSSDNet model, for frequencies up to 4~kHz. Here we see which spectral bins lend the most support on average to either bona fide or spoofed classes.  In addition to other less substantial differences, there is predominantly greater support for the bona fide class at ~0.5~kHz but substantially greater support for the spoofed class at ~0.6~kHz. We observed many instances of such differences suggesting that some spoofing attacks leave artefacts in specific spectral intervals while they are largely effective in replicating the characteristics of bona fide speech in others.  Similar observations have been reported previously~\cite{tak20_odyssey, zhang21da_cas}.  Such characteristics may help not just to distinguish between bona fide and spoofed speech, but also to identity a particular spoofing attack algorithm or its nature, e.g., whether it is a synthetic speech, converted voice or replay attack.

\subsection{Classifier differences}
\begin{figure}[!t]
     \centering
     \begin{subfigure}[b]{0.45\textwidth}
         \centering
        \includegraphics[width=\textwidth]{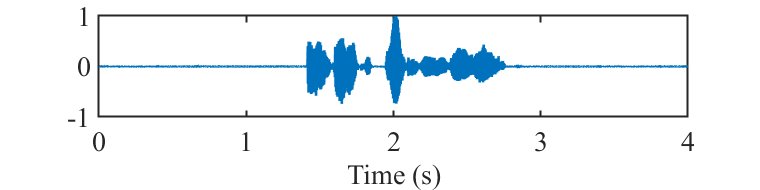}
         \caption{waveform}
         \label{fig:waveform_4428024}
     \end{subfigure}
     \begin{subfigure}[b]{0.45\textwidth}
         \centering
        \includegraphics[width=\textwidth]{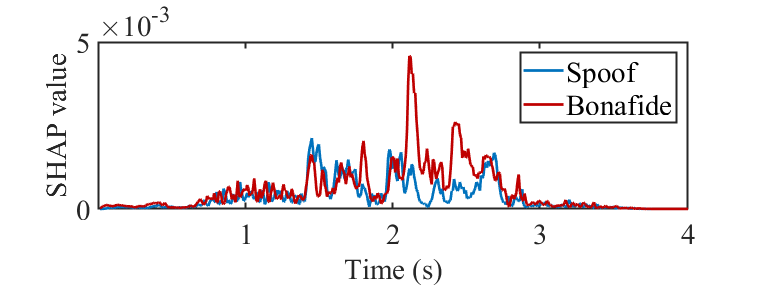} \caption{2D-Res-TSSDNet}
         \label{fig:tssdnet_4428024}
     \end{subfigure}
     \begin{subfigure}[b]{0.45\textwidth}
         \centering
        \includegraphics[width=\textwidth]{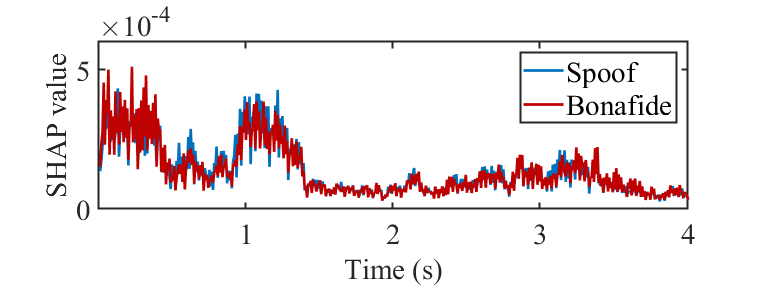} \caption{PC-DARTS}
         \label{fig:pcdarts_4428024}
     \end{subfigure}
        \caption{SHAP values for the `LA\_E\_4428024' utterance and both the 2D-Res-TSSDNet and PC-DARTS models.} 
        \label{fig:classifiers}
\end{figure}
    Fig.~\ref{fig:classifiers} shows an example for which SHAP analysis reveals differences in classifier behaviour.  The two plots show frequency-averaged SHAP values against time for the 2D-Res-TSSDNet classifier (middle) and the PC-DARTS classifier (bottom) and in both cases the support for the spoof class (blue) and bona fide class (red).  The classifiers are shown to apply attention to different intervals.  The PC-DARTS model derives its output mostly from non-speech segments whereas the 2D-Res-TSSDNet model applies greater attention to speech intervals.  We observed other, more subtle differences in classifier behaviour. SHAP values for the PC-DARTS model are notably more noisy than those of 2D-Res-TSSDNet model.  We believe that these differences stem from the relative simplicity of the PC-DARTS model which contains only 0.1~M network parameters, 0.87~M fewer than the 2D-Res-TSSDNet model, possibly implying that the former has insufficient learning capacity, leading to noisier outputs.  These characteristics might also be caused by the high number of dilated convolutions in the learned architecture.  We observed more stable SHAP values when dilated convolution operations are excluded from the architecture search space.  SHAP analysis can hence also be used to explore lower-level classifier behaviours. Instances where different classifiers exhibit different behaviour might also help to improve performance in cases where single system solutions are preferred to system fusion. 

\section{Conclusions}

    This paper demonstrates how DeepSHAP can be applied to explain what influences the outputs produced by a spoofing detection model.  The examples shown in the paper show how SHAP analysis can be used to highlight the attention applied by a given classifier at low-level spectro-temporal intervals.  Nonetheless, the tool offers the basis for what is needed to explore higher-level explanations.  It will be of interest to explore, e.g., whether we can make the link between SHAP results, classifier outputs and specific speech units or spoofing attack algorithms (e.g., synthetic speech, converted voice and replay) and their algorithmic components (e.g., waveform models, recording devices and microphones).  Other future directions include the use of SHAP to explore classifier differences in an attempt to explain why some classifiers perform better than others, and under which conditions.  In this context, it will be of interest to develop SHAP-based approaches that can be used to compare systems that operate upon hand-crafted spectro-temporal decompositions to those that operate directly upon raw waveforms. Ultimately, of course, the goal is to exploit what we can learned from SHAP analysis to design better performing, more reliable models.

\balance
\bibliographystyle{IEEEbib}
\bibliography{references}

\end{document}